\documentclass{article}

\usepackage{tikz}
\usepackage{subfigure}
\usetikzlibrary{calc}

\usepackage{amsmath}
\usepackage{amssymb}
\usepackage{amsfonts}
\usepackage{esint}
\usepackage{mathrsfs}
\usepackage{amsthm}
\usepackage{amscd}
\usepackage{longtable}

\usepackage{listings}
\usepackage{booktabs,longtable}
\usepackage{graphicx}

\usepackage{multirow}

\newcommand{\ppulp}[1]{{\mathscr #1}}

\newcommand{\ppulpm}[1]{\pmb{\ppulp{#1}}}

\newcommand{\transp}[1]{#1^{\rm T}}

\DeclareMathOperator{\ber}{ber}
\DeclareMathOperator{\bei}{bei}
\newcommand{\vect}[1]{\mathbf{#1}}
\newcommand{\ver}[1]{\hat{#1}}
\newcommand{\matr}[1]{\mathbf{#1}}
\newcommand{\abs}[1]{\left \lvert #1 \right\rvert }

\renewcommand{\Re}[1]{\mathbb{R}\mathrm{e}\left \{#1\right\} }
\renewcommand{\Im}[1]{\mathbb{I}\mathrm{m}\left \{#1\right\} }

\newcommand{\pref}[1]{(\ref{#1})}
\newcommand{\junk}[1] {}

\def\XXint#1#2#3{{\setbox0=\hbox{$#1{#2#3}{\int}$}
\vcenter{\hbox{$#2#3$}}\kern-.5\wd0}}

\begin{document}

\title{Fast Computation of the Series Impedance of Power Cables with Inclusion of Skin and Proximity Effects}

\author{Utkarsh~R.~Patel,
        Bj\o rn~Gustavsen,
        and~Piero~Triverio
\thanks{This work was partially supported by the Norwegian Research Council (RENERGI Programme) with additional support from an industry consortium led by SINTEF Energy Research.}
\thanks{U.~R.~Patel and P.~Triverio are with the Edward S. Rogers Sr. Department of Electrical and Computer Engineering, University of Toronto, Toronto, M5S 3G4 Canada (email: utkarsh.patel@mail.utoronto.ca, piero.triverio@utoronto.ca).}
\thanks{B.~Gustavsen is with SINTEF Energy Research, Trondheim N-7465, Norway (e-mail: bjorn.gustavsen@sintef.no).}
\\ Published in {\em IEEE Transactions on Power Delivery}, vol. 28, no. 4, \\
Oct. 2013. DOI: 10.1109/TPWRD.2013.2267098.
\thanks{Published with the revised-title ``An Equivalent Surface Current Approach for the Computation of the Series Impedance of Power Cables with Inclusion of Skin and Proximity Effects''}
}

\maketitle

\begin{abstract}
We present an efficient numerical technique for calculating the series impedance matrix of systems with round conductors. The method is based on a surface admittance operator in combination with the method of moments and it accurately predicts both skin and proximity effects. Application to a three-phase armored cable with wire screens demonstrates a speed-up by a factor of about 100 compared to a finite elements computation. The inclusion of proximity effect in combination with the high efficiency makes the new method very attractive for cable modeling within EMTP-type simulation tools. Currently, these tools can only take skin effect into account.   
\end{abstract}

\section{Introduction and Motivation}

Electromagnetic transients have a significant impact on the design, operation and performance of electrical power systems~\cite{Cho96,Wat03}. Transients can be caused by several phenomena such as lightning discharges, breaker operations, faults, and the use of power electronics converters. Since electromagnetic transients involve a wide band of frequencies, ranging from DC up to the low MHz range, their simulation requires broadband and accurate models for all network components to predict the network response outside the nominal sinusoidal regime. For the modeling of underground cables, it is necessary to calculate the per-unit-length (p.u.l.) cable series impedance matrix~\cite{Pau07} over a wide band of discrete frequencies  while taking into account frequency-dependent phenomena such as skin and proximity effects. The series impedance is next used as input data for alternative frequency-dependent cable models \cite{LMarti82},\cite{Mor99}.

Traditionally, the series impedance is calculated using analytic formulas which account for skin effect only, since they assume a symmetrical distribution of current in all conductors~\cite{Pau07}. These formulas are combined with systematic procedures for computing the series impedance matrix for multi-conductor systems~\cite{Ame80}. Although simple and highly efficient, this traditional approach neglects proximity effects. Proximity-aware formulas are available only for two conductor systems~\cite{Car21,Pau07}. Ignoring proximity effect is acceptable for overhead lines and for widely spaced single core cable systems. However, in the case of three-phase cables, pipe-type cables and closely packed single core cables, the small distance between conductors in combination with the non-coaxial arrangement leads to significant proximity effects. This issue is relevant also in the modeling of umbilical cables for offshore oil and gas power supply and control~\cite{B09}. Ignoring proximity effect leads to an underestimate of the cable losses at the operating frequency, and the transient waveforms are also affected, in particular in situations where waves propagate between the screens and between the screens and ground~\cite{Bjorn95,Unnur2011}. The latter situation is highly relevant in the simulation of cross-bonded cable systems.

Because of the limitations of analytic formulas, several numerical techniques have been proposed. The harmonic expansion method accounts for proximity effects at high frequencies under the assumption that the skin effect is fully developed~\cite{Pau07,Cle75,Bra08,Sav95}. Techniques based on the finite element method (FEM)~\cite{Wei82,B09,Cri89} fully predict proximity effect at both low and high frequencies, but they tend to be excessively time-consuming because of the fine mesh required to properly discretize the cross section. A similar issue arises with techniques based on conductor partitioning~\cite{Ame92,Com73,Dea87,Pag12,Riv02}. Moreover, with both FEM and conductor partitioning techniques, the discretization must be refined as frequency increases to properly capture the pronounced skin effect, further reducing computational efficiency.

In this paper, we overcome these issues by developing an efficient numerical technique for computing the series impedance matrix of cables with round conductors. Our approach, denoted henceforth as MoM-SO, combines the method of moments~(MoM) with a surface admittance operator~(SO) introduced in~\cite{DeZ05} for rectangular conductors. The proposed approach is not available elsewhere in the literature, since the authors of~\cite{DeZ05}, after presenting the surface admittance operator for both rectangular and round conductors, focus their attention on the first case. Through the surface operator, we replace each conductor with the surrounding medium, while introducing an equivalent current density on the surface of each conductor. The current density and the longitudinal electric field on all conductors are then related through the electric field integral equation~\cite{Bal05}, and their spatial dependence is expressed in terms of Fourier components. Finally, the method of moments~\cite{Wal08} is applied to compute the p.u.l. parameters of the line. Numerical results will show that the proposed approach is much more efficient than state-of-the-art FEM techniques, since few Fourier components are sufficient to accurately model skin and proximity effect at any frequency. Moreover, the discretization of the electric field integral equation is entirely perfomed using analytic formulas, avoiding numerical integration used in previous works~\cite{Sav95,DeZ05}. This achievement further improves the robustness and speed of MoM-SO.

The paper is organized as follows. In Sec.~\ref{sec:theory} we state the problem from a theoretical standpoint and review the two fundamental relations exploited in this work, namely the electric field integral equation and the surface admittance operator. In Sec.~\ref{sec:num}, the two relations are discretized with the method of moments. In Sec.~\ref{sec:results}, the new MoM-SO method is first validated against analytical formulas for a simple two-conductor system, and then compared against FEM on a three-phase armored cable.

\section{Impedance Computation via a Surface Admittance Operator}
\label{sec:theory}

\subsection{Problem Statement}
We consider a transmission line made by $P$ round conductors parallel to the $z$ axis and surrounded by a homogeneous medium. An example of line cross section is shown in Fig.~\ref{fig:crosssection}, where the permittivity, permeability and conductivity of the conductors are denoted as $\varepsilon$, $\mu$, and $\sigma$ respectively.
\begin{figure}[!t]
\centering
\begin{tikzpicture}
\draw [lightgray,fill=lightgray] (0,0) rectangle (6,2);
\node [above] at (5,0) {$\varepsilon_{out}, \mu_0$};

\draw [black,fill=white] (1.7,1) circle [radius=.8] node [below] {$\varepsilon, \mu, \sigma$};
\node [right] at (2.8,1) {$\ver{n}$};

\draw [black,fill=white] (4.5,1.2) circle [radius=.6];

\draw [thick, ->] (2.5,1) -- (2.8,1);

\end{tikzpicture}
\caption{Sample line cross section. The conductors are depicted in white (permittivity: $\varepsilon$, permeability: $\mu$, conductivity: $\sigma$). The surrounding medium is depicted in grey (permittivity: $\varepsilon_{out}$, permeability: $\mu_0$). The unit vector $\ver{n}$ is the normal to the conductors surface.}
\label{fig:crosssection}
\vspace{0.5cm}

\begin{tikzpicture}
\draw [lightgray,fill=lightgray] (0,0) rectangle (6,2);
\node [above] at (5,0) {$\varepsilon_{out}, \mu_0$};

\draw [black, dashed] (1.7,1) circle [radius=.8];
\node [right] at (.7,.3) {$c_p$};
\draw [dotted] (1.7,1) -- (2.8,1);
\draw [dotted] (1.7,1) -- +(45:1.1);
\draw [thick, ->] (1.7,1) -- +(45:.8);
\node [above] at (1.8,1.1) {$a_p$};
\draw [->] (2.8,1) arc (0:45:1.1);
\node [above] at (2.9,1.3) {$\theta$};

\foreach \a in {20, 80,...,320} {
      \draw ({1.7+.8*cos(\a)},{1+.8*sin(\a)}) circle [radius=.1];
      \draw [fill=black] ({1.7+.8*cos(\a)},{1+.8*sin(\a)}) circle [radius=.04];
}

\node [above left] at (1,1.4) {$J_s$};

\draw [black, dashed] (4.5,1.2) circle [radius=.6];
\foreach \a in {20, 80,...,320} {
      \draw ({4.5+.6*cos(\a)},{1.2+.6*sin(\a)}) circle [radius=.1];
      \draw [fill=black] ({4.5+.6*cos(\a)},{1.2+.6*sin(\a)}) circle [radius=.04];
}

\end{tikzpicture}
\caption{Line cross section after application of the equivalence theorem. The conductors medium is replaced by the surrounding medium, and an equivalent current density $J_s$ is introduced on the surface of the conductors. The contour and radius of the $p$-th conductor are denoted with $c_p$ and $a_p$, respectively.}
\label{fig:crosssection_equivalence}

\end{figure}
The surrounding medium is assumed to be lossless, with permittivity $\varepsilon_{out}$ and permeability $\mu_0$.  Our goal is to compute the per-unit-length (p.u.l.) resistance $\ppulpm{R}(\omega)$ and inductance $\ppulpm{L}(\omega)$ matrices of the line as defined by the Telegraphers' equation~\cite{Pau07}
\begin{equation}
\frac{\partial \vect{V}}{\partial z} = - \left [\ppulpm{R}(\omega) + j\omega\ppulpm{L}(\omega)\right ] \vect{I}\,
\label{eq:pteleg}
\end{equation}
where $\vect{V} = [V_1 \cdots V_P]^T$ is a $P \times 1$ vector collecting the potential $V_p$ of each conductor. Similarly, the $P \times 1$ vector $\vect{I} = [I_1 \cdots I_P]^T$ is formed by the current $I_p$ in each conductor. The parameters $\ppulpm{R}(\omega)$ and $\ppulpm{L}(\omega)$ in~\pref{eq:pteleg} are commonly referred as \emph{partial} p.u.l. parameters~\cite{Pau10}. From them, one can easily obtain the p.u.l. parameters of the line with any conductor taken as reference for the voltages and as return path for the currents~\cite{Pau07}. In this work, the cable parameters are computed assuming that the electric and magnetic fields are longitudinally invariant along the cable. We neglect ``end effects'' that may arise from the finite length of the cable, and which may be noticeable for short-length cables~\cite{Ame93}. We refer the Reader to~\cite{Ame05,Ame93} for more details on this aspect.

\subsection{Surface Admittance Operator}
\label{sec:surfaceadmittance}

In order to compute the p.u.l. impedance of the line, we follow the approach of~\cite{DeZ05} which relies on a surface admittance operator. We replace each conductor with the surrounding medium and, to maintain the electric field $E_{z}$ \emph{outside} the conductors' volume unchanged, we introduce a surface current density $J_s$ on their contour, as shown in Fig.~\ref{fig:crosssection_equivalence}. The electric field \emph{inside} the conductors' volume takes instead a fictitious value $\widetilde{E}_{z}$. The current density $J_s$, directed along $z$, can be found with the equivalence theorem~\cite{Har61,DeZ05} and reads
\begin{equation}
{J_s} = H_t - \widetilde{H}_{t}\,,
\label{eq:bound1}
\end{equation}
where $H_t$ is the component of the magnetic field tangential to the conductor's surface, evaluated \emph{before} the application of the equivalence theorem (configuration of Fig.~\ref{fig:crosssection}). $\widetilde{H}_{t}$ is the same quantity evaluated \emph{after} the equivalence theorem has been applied (configuration of Fig.~\ref{fig:crosssection_equivalence}).

On the boundary of each conductor, the tangential component of the magnetic field is related to the longitudinal electric field~\cite{DeZ05}
\begin{equation}
H_{t} = \frac{1}{j\omega \mu} \frac{\partial  E_z}{\partial n}\,,
\label{eq:HErel}
\end{equation}
where $\frac{\partial}{\partial n}$ denotes the directional derivative~\cite{Kap91} with respect to the unit vector $\ver{n}$ normal to the conductors surface. Similarly, for $\widetilde{H}_{t}$ we can write
\begin{equation}
\widetilde{H}_{t} = \frac{1}{j\omega \mu_0} \frac{\partial \widetilde{E}_{z}}{\partial n}\,.
\label{eq:HErel2}
\end{equation}
By substituting~\pref{eq:HErel} and~\pref{eq:HErel2} into~\pref{eq:bound1}, we obtain the relation
\begin{equation}
{J_s} = \frac{1}{j\omega} \left [ \frac{1}{\mu} \frac{\partial E_z}{\partial n} - \frac{1}{\mu_0} \frac{\partial \widetilde{E}_{z}}{\partial n} \right ] \,,
\label{eq:bound2}
\end{equation}
which defines a surface admittance operator that maps the electric field on the conductor's boundary onto the equivalent current density $J_s$. Equation~\pref{eq:bound2} extends the formula given in~\cite{DeZ05} to the case of magnetic conductors ($\mu \neq \mu_0$).

\subsection{Electric Field Integral Equation}

After applying the equivalence theorem to each conductor, the medium becomes homogeneous, and we can easily relate the equivalent current density $J_s$ to the electric field by means of the electric field integral equation~\cite{Bal05}. This process involves only the boundary $c_p$ of each conductor, that we describe with the position vector
\begin{equation}
\vec{r}_p(\theta) = 	\left (x_p + a_p \cos \theta  \right ) { \ver{x}} +
			\left (y_p + a_p \sin \theta  \right ) { \ver{y}}\,,
\label{eq:contour}
\end{equation}
where $\theta$ is the azimuthal coordinate, $(x_p,y_p)$ are the coordinates of the conductor center, and $a_p$ is the conductor radius, as illustrated in Fig.~\ref{fig:crosssection_equivalence}. The unit vectors $\ver{x}$ and $\ver{y}$ are aligned with the $x$ and $y$ axis, respectively. We denote the equivalent current density and the electric field on the boundary of the $p$-th conductor as $J_s^{(p)}(\theta)$ and $E_z^{(p)}(\theta)$, respectively.
Using the electric field integral equation, we can express the electric field $E_z^{(p)}(\theta)$ on the surface of the $p$-th conductor as 
\begin{multline}
E^{(p)}_z(\theta) =  j \omega \mu_0 \sum_{q=1}^{P} \int_0^{2\pi} J^{(q)}_{s}(\theta') G \left ( \vec{r}_p(\theta),\vec{r}_q(\theta') \right ) a_q d\theta' \\ - \frac{\partial V}{\partial z}  \,,
\label{eq:efie}
\end{multline}
where the first term accounts for the field generated by the current on each conductor, while the second term is related to the scalar potential $V$. The integral kernel
\begin{equation}
G(\vec{r}_p,\vec{r}_q) = \frac{1}{2\pi} \ln \abs{\vec{r}_p-\vec{r}_q}
\label{eq:green}
\end{equation}
 is the Green's function of an infinite space~\cite{Har61}. On the $p$-th conductor, the scalar potential $V$ is equal to the conductor potential $V_p$ that appears in the Telegraphers' equation~\pref{eq:pteleg}. Therefore, we can replace the last term in~\pref{eq:efie} with~\pref{eq:pteleg}, obtaining
\begin{align}
E_z^{(p)}(\theta) &= j \omega \mu_0 
\sum_{q=1}^{P} \int_{0}^{2\pi} J^{(q)}_{s}(\theta') G \left (\vec{r}_p(\theta),\vec{r}_q(\theta') \right ) a_q d\theta' \nonumber \\
&+ \sum_{q=1}^P
\left [ \ppulpm{R}_{pq}(\omega)+j\omega \ppulpm{L}_{pq}(\omega) \right ] I_q\,, 
\label{eq:efie3}
\end{align}
where $\ppulpm{R}_{pq}(\omega)$ and $\ppulpm{L}_{pq}(\omega)$ are the elements in position $(p,q)$ of the matrices $\ppulpm{R}(\omega)$ and $\ppulpm{L}(\omega)$, respectively. 
Equation~\pref{eq:efie3} combined with the surface admittance operator~\pref{eq:bound2} will allow us to compute the p.u.l. parameters of the line. In the next Section, we introduce a discretization of these relations suitable for numerical computations.

\section{Numerical Formulation}
\label{sec:num}

\subsection{Discretization of the Surface Admittance Operator}

Given the cylindrical geometry of the conductors, we approximate the field and the current on the $p$-th conductor by means of a truncated Fourier series
\begin{eqnarray}
E^{(p)}_z(\theta) & = & \sum_{n=-N_p}^{N_p} E^{(p)}_{n} e^{j n \theta}\,,
\label{eq:Eseries} \\
J^{(p)}_s(\theta) & = & \frac{1}{2\pi a_p} \sum_{n=-N_p}^{N_p} J^{(p)}_{n} e^{j n \theta}\,. 
\label{eq:Jseries}
\end{eqnarray}
The truncation order $N_p$ controls the accuracy and the computational cost of the numerical technique. Examples will show that a low $N_p$, of the order of $2-3$, delivers very accurate results while minimizing the computation time.
Owing to the normalization factor $\frac{1}{2\pi a_p}$, the total current $I_p$ flowing in the $p$-th conductor is simply given by the constant term of the series~\cite{DeZ05} 
\begin{equation}
I_p = J^{(p)}_{0} \qquad p=1,\dots,P\,.
\label{eq:Ip}
\end{equation}
When $E^{(p)}_z$ and $J^{(p)}_s$ are expressed in Fourier series, the surface admittance operator~\pref{eq:bound2} can be rewritten in terms of the Fourier coefficients as~\cite{DeZ05}
\begin{equation}
J^{(p)}_{n} = E^{(p)}_{n} \frac{2\pi}{j\omega } \biggl [ \frac{k a_p {\cal J}_{|n|}'(ka_p)}{\mu {\cal J}_{|n|}(ka_p)} - \frac{k_{out} a_p {\cal J}_{|n|}'(k_{out}a_p)}{\mu_o {\cal J}_{|n|}(k_{out}a_p)}  \biggl ]\,,
\label{eq:JCoeff}
\end{equation}
where ${\cal J}_{|n|}(.)$ is the Bessel function of the first kind~\cite{Abr64} of order $|n|$, and ${\cal J}_{|n|}'(.)$ is its derivative. The wavenumber in the conductors and in the surrounding medium are given respectively by
\begin{align}
k & = \sqrt{ \omega \mu (\omega \varepsilon -j \sigma)}\,, \\
k_{out} & = \omega \sqrt{\mu_0 \varepsilon_{out}}\,.
\label{eq:k}
\end{align}
In order to simplify the oncoming equations, we introduce a compact matrix notation. We collect all field coefficients $E^{(p)}_{n}$ in the column vector
\begin{equation}
	\vect{E} = 
	\begin{bmatrix}
		E^{(1)}_{-N_1} & \cdots & E^{(1)}_{N_1} & E^{(2)}_{-N_2} & \cdots & E^{(2)}_{N_2} & \cdots
	\end{bmatrix}^T\,,
	\label{eq:Evect}
\end{equation}
and all current coefficients $J^{(p)}_{n}$ in
\begin{equation}
	\vect{J} = 
	\begin{bmatrix}
		J^{(1)}_{-N_1} & \cdots & J^{(1)}_{N_1} & J^{(2)}_{-N_2} & \cdots & J^{(2)}_{N_2} & \cdots
	\end{bmatrix}^T\,.
	\label{eq:Jvect}
\end{equation}
The vectors $\vect{E}$ and $\vect{J}$ have size
\begin{equation}
	N = \sum_{p=1}^P (2 N_p +1)\,,
	\label{eq:N}
\end{equation}
the total number of field and current coefficients.
Relation~\pref{eq:JCoeff} can be written in terms of $\vect{E}$ and $\vect{J}$ as
\begin{equation}
\vect{J} = \matr{Y}_{s} \vect{E}\,.
\label{eq:sad}
\end{equation}
where $\matr{Y}_{s}$ is a diagonal matrix. This matrix is the discrete version of the surface admittance operator defined by~\pref{eq:bound2} for each round conductor. Finally,~\pref{eq:Ip} can be written in terms of $\vect{I}$ and $\vect{J}$ as
\begin{equation}
\vect{I} = \matr{U}^T\vect{J}\,,
\label{eq:UJrel}
\end{equation}
where $\matr{U}$ is a constant $N \times P$ matrix made by all zeros and a single ``1'' in each column. In column $p$, the ``1'' is in the same row as the coefficient $J_0^{(p)}$ in~\pref{eq:Jvect}.

\subsection{Discretization of the Electric Field Integral Equation}

We now cast the electric field integral equation~\pref{eq:efie3} into a set of algebraic equations using the method of moments, a numerical method to solve integral and differential equations~\cite{Wal08}. First, we substitute~\pref{eq:Eseries} and~\pref{eq:Jseries} into~\pref{eq:efie3}, obtaining
\begin{align}
\sum_{n=-N_p}^{N_p} & E^{(p)}_{n} e^{j n \theta} = \nonumber \\
& \frac{j \omega \mu_0}{2 \pi} \sum_{q=1}^{P} 
\sum_{n=-N_q}^{N_q} J^{(q)}_{n} \int_{0}^{2\pi} e^{j n \theta'} G \left (\vec{r}_p(\theta),\vec{r}_q(\theta') \right ) d \theta' \nonumber \\
&+ \sum_{q=1}^P \left [ \ppulpm{R}_{pq}(\omega)+j\omega \ppulpm{L}_{pq}(\omega) \right ] I_q\,.
\label{eq:mom1}
\end{align}
Then, we project~\pref{eq:mom1} onto the Fourier basis functions $e^{j n' \theta}$ by applying the operator
\begin{equation}
\int_0^{2\pi} [.] e^{-j n' \theta} d\theta \qquad n' = -N_p,\dots, N_p
\label{eq:galerk}
\end{equation}
to both sides of the equation, obtaining
\begin{multline}
E^{(p)}_{n'} =  j \omega \mu_0 \sum_{q=1}^{P} \sum_{n=-N_q}^{N_q} \matr{G}^{(p,q)}_{n'n} J^{(q)}_{n} + \\
\delta_{n',0} \sum_{q=1}^P \left [ \ppulpm{R}_{pq}(\omega)+j\omega \ppulpm{L}_{pq}(\omega) \right ] I_q\,,
\label{eq:mom4}
\end{multline}
for $p=1,\dots,P$, where
\begin{equation}
\delta_{n',0} = \begin{cases}
	1 & \text{when } n' = 0 \\
	0 & \text {when } n' \neq 0\,,
\end{cases}
\end{equation}
and where $\matr{G}^{(p,q)}_{n'n}$ denotes the $(n',n)$ entry of the matrix $\matr{G}^{(p,q)}$. This matrix describes the contribution of the current on the $q$-th conductor to the field on the $p$-th conductor. The entries of $\matr{G}^{(p,q)}$ are given by the double integral
\begin{equation}
\matr{G}^{(p,q)}_{n'n}=\frac{1}{(2 \pi)^2} \int_{0}^{2\pi}\int_{0}^{2\pi} 
G \left (\vec{r}_p(\theta),\vec{r}_q(\theta') \right )
e^{j (n \theta' - n' \theta)} d\theta d\theta' \,,
\label{eq:greenentry}
\end{equation}
which can be computed analytically as shown in the Appendix. Using the matrix notation set in~\pref{eq:Evect} and~\pref{eq:Jvect}, the equations~\pref{eq:mom4} can be written in compact form as
\begin{equation}
\vect{E} = j\omega \mu_0 \matr{G} \vect{J} + \matr{U} \left [\ppulpm{R}(\omega) + j \omega\ppulpm{L}(\omega)\right ] \vect{I}\,,
\label{eq:mom}
\end{equation}
where $\matr{G}$ is the block matrix
\begin{equation}
	\matr{G} = 
	\begin{bmatrix}
		\matr{G}^{(1,1)} & \cdots & \matr{G}^{(1,P)}\\
		\vdots & \ddots & \vdots	\\		
		\matr{G}^{(P,1)} & \cdots & \matr{G}^{(P,P)}				
	\end{bmatrix}\,.
	\label{eq:G}
\end{equation}
Equation~\pref{eq:mom} is the discrete counterpart of the electric field integral equation~\pref{eq:efie3}.

\subsection{Computation of the Per-Unit-Length Parameters}

The p.u.l. parameters of the line can be obtained by combining the discretized surface admittance operator~\pref{eq:sad} with the discretized electric field integral equation~\pref{eq:mom} as follows. First, we left multiply~\pref{eq:mom} by $\matr{Y}_s$ and, using~\pref{eq:sad}, we obtain
\begin{equation}
\vect{J} = j\omega \mu_0 \matr{Y}_s \matr{G} \vect{J} + \matr{Y}_s \matr{U} \left [\ppulpm{R}(\omega) + j \omega\ppulpm{L}(\omega)\right ] \vect{I}\,.
\label{eq:ppulp1}
\end{equation}
The current density coefficients $\vect{J}$ can be expressed as
\begin{equation}
\vect{J} = (\matr{1} - j\omega \mu_0 \matr{Y}_s \matr{G})^{-1} \matr{Y}_s \matr{U} \left [\ppulpm{R}(\omega) + j \omega\ppulpm{L}(\omega)\right ] \vect{I}\,,
\label{eq:ppulp2}
\end{equation}
where $\matr{1}$ denotes the $N \times N$ identity matrix. Left multiplication of~\pref{eq:ppulp2} by $\transp{\matr{U}}$ leads to the following expression for the line currents $\vect{I}$
\begin{equation}
\vect{I} =  \left [ \transp{\matr{U}} (\matr{1} - j\omega \mu_0 \matr{Y}_s \matr{G})^{-1} \matr{Y}_s \matr{U} \right ] \cdot \left [ \ppulpm{R}(\omega) + j \omega\ppulpm{L}(\omega)\right ] \vect{I}\,.
\label{eq:ppulp3}
\end{equation}
Since~\pref{eq:ppulp3} must hold for any $\vect{I}$, the product of the two expressions inside the square brackets must be equal to the identity matrix. Consequently, we have that
\begin{equation}
\ppulpm{R}(\omega) + j \omega\ppulpm{L}(\omega) = \left [ \transp{\matr{U}} (\matr{1} - j\omega \mu_0 \matr{Y}_s \matr{G})^{-1} \matr{Y}_s \matr{U} \right ]^{-1}\,.
\label{eq:ppulp4}
\end{equation}
By taking the real and imaginary part of~\pref{eq:ppulp4} we finally obtain the formulas for computing the p.u.l. resistance and inductance matrices
\begin{align}
\ppulpm{R}(\omega) & = \Re{\left [\transp{\matr{U}} (\matr{1} - j\omega \mu_0 \matr{Y}_s \matr{G})^{-1} \matr{Y}_s \matr{U} \right ]^{-1}}\,,
\label{eq:ppulpR} \\
\ppulpm{L}(\omega) & = \omega^{-1} \Im{\left [\transp{\matr{U}} (\matr{1} - j\omega \mu_0 \matr{Y}_s \matr{G})^{-1} \matr{Y}_s \matr{U} \right ]^{-1}}\,.
\label{eq:ppulpL}
\end{align}

\section{Numerical results}
\label{sec:results}

\subsection{Two Round Conductors}
\label{sec:2cond}

In order to validate the proposed technique against analytic formulas, we consider a line made by two parallel round conductors with radius $a = 10$~mm made of copper ($\sigma=58 \cdot 10^6$~S/m, $\mu = \mu_0$). Two different values for the center-to-center distance between the wires have been used, namely $D = 100$~mm and $D = 25$~mm. In the first case proximity effect is negligible, due to the wide separation. In the second case it is instead significant. 

\begin{figure}[t]
\centering
\includegraphics[width=.75\columnwidth, viewport= 150 220 470 540]{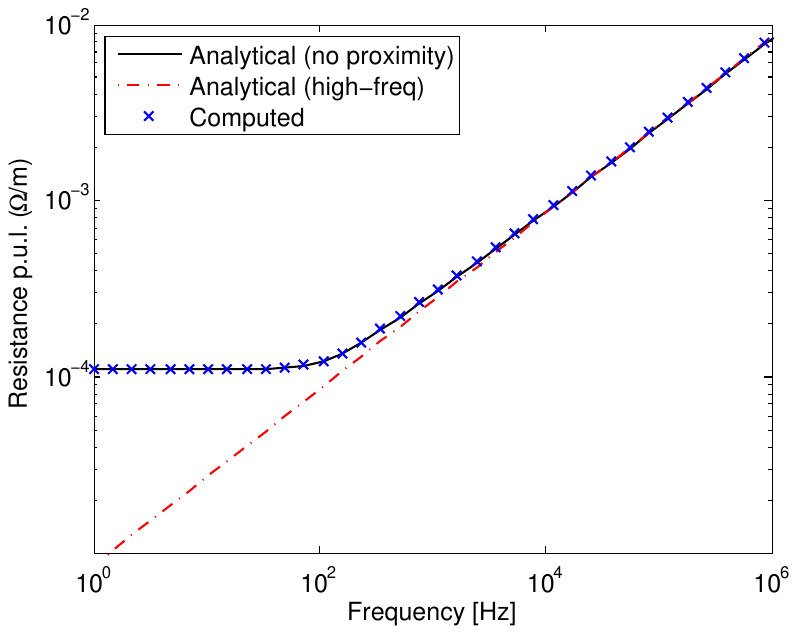}
\includegraphics[width=.75\columnwidth, viewport= 150 240 470 540]{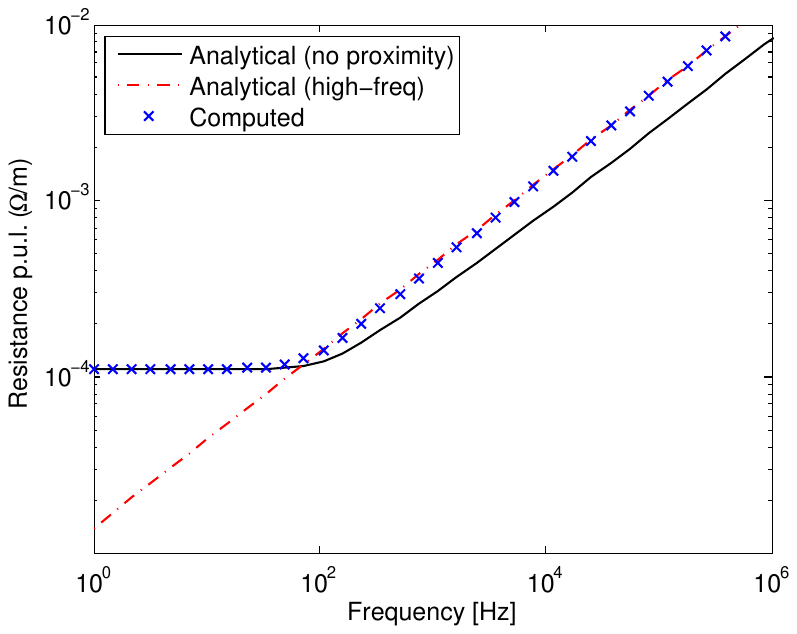}
\caption{P.u.l. resistance of the two wires line of Sec.~\ref{sec:2cond} computed with MoM-SO (crosses), the analytic formula~\pref{eq:R} valid at high frequency (dash-dot line), and formula~\pref{eq:Zfreqdep} (solid line).
Wires separation is 100~mm (top panel) and 25~mm (bottom panel).}
\label{fig:res1}
\end{figure}

The p.u.l. computed with a MATLAB implementation of MoM-SO have been compared against two different sets of analytic formulas. The first set is valid at high frequency~\cite{Pau07} because it assumes a fully-developed skin effect, and gives the p.u.l. resistance and inductance as
\begin{align}
\ppulp{R} & = \frac{R_s}{\pi a} \frac{\frac{D}{2a}}{\sqrt{ \left (\frac{D}{2a} \right )^2 -1}}\, ,  \label{eq:R} \\
\ppulp{L}_{ext} & = \frac{\mu_0}{\pi} \cosh^{-1} \left ( \frac{D}{2a} \right ) \label{eq:Lext} \, ,
\end{align}
where $R_s = (\sigma \delta)^{-1} $ is the surface resistance and
\begin{equation}
	\delta = \frac{1}{\sqrt{\pi f \mu_0 \sigma}}
\end{equation}
is the skin depth.
The second set of formulas accounts for the frequency dependence of the internal impedance of the wire $\ppulp{Z}_{int}$, which can be calculated analytically under the assumption of wide separation~\cite{Pau07}
\begin{equation}
	\ppulp{Z}_{int} = \frac{1}{\sqrt{2} \pi a \sigma \delta} \frac{\ber(\xi) + j \bei (\xi)}{\bei'(\xi) - j \ber' (\xi)}\,,
	\label{eq:Zint}	
\end{equation}
where $\xi = \sqrt{2} \frac{a}{\delta}$. The Kelvin functions $\ber(\xi)$ and $\bei(\xi)$ are the real and imaginary part of ${\cal J}_0(\xi e^{j \frac{3}{4} \pi})$, respectively~\cite{Abr64}.
The total p.u.l. impedance of the line is thus
\begin{equation}
	\ppulp{Z} = 2 \ppulp{Z}_{int} + j \omega \ppulp{L}_{ext}\,.
	\label{eq:Zfreqdep}
\end{equation}
Formulas~\pref{eq:R}-\pref{eq:Lext} account for proximity effect, which is instead neglected in~\pref{eq:Zint}. 

\begin{figure}[t]
\centering
\includegraphics[width=.75\columnwidth, viewport= 150 220 470 540]{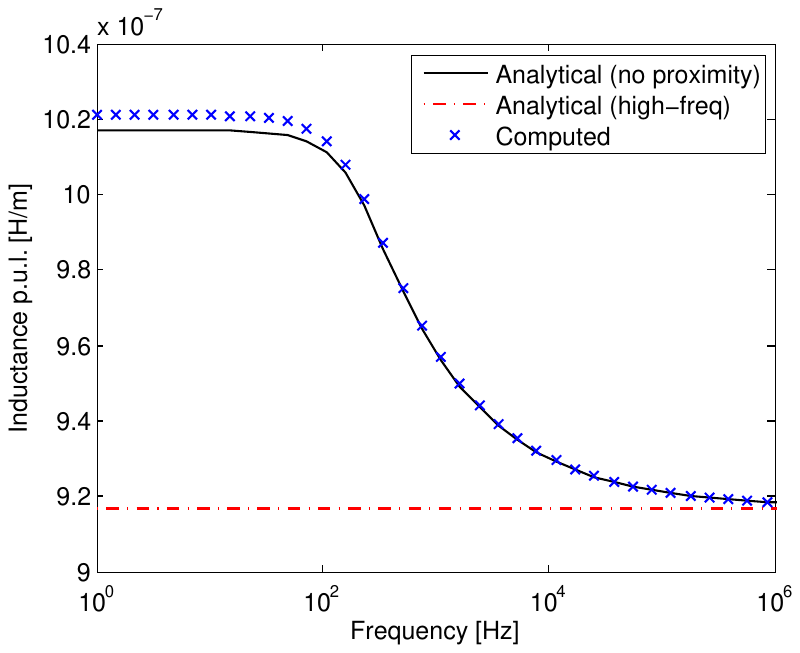}
\includegraphics[width=.75\columnwidth, viewport= 150 240 470 540]{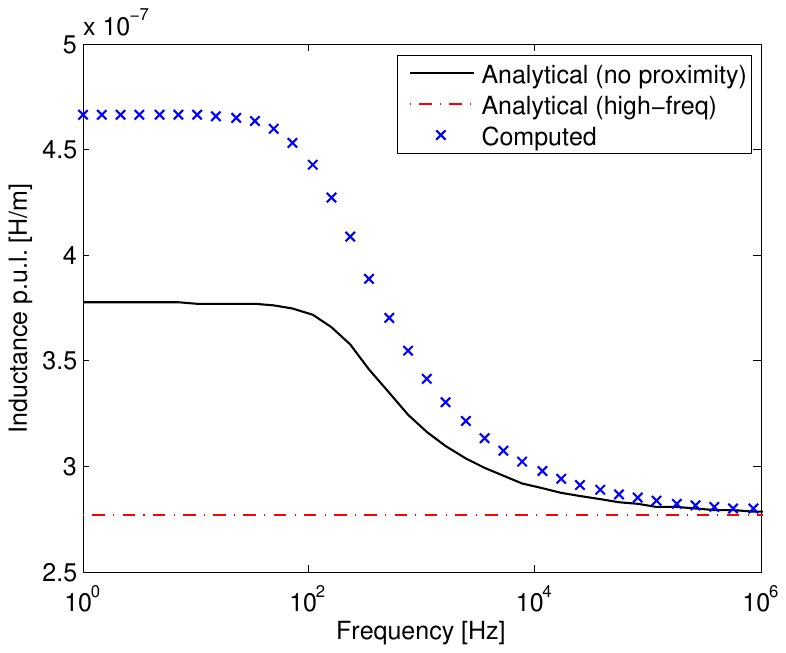}
\caption{P.u.l. inductance of the two wires line of Sec.~\ref{sec:2cond} computed with MoM-SO (crosses), the analytic formula~\pref{eq:Lext} valid at high frequency (dash-dot line), and formula~\pref{eq:Zfreqdep} (solid line).
Wires separation is 100~mm (top panel) and 25~mm (bottom panel).}
\label{fig:ind1}
\end{figure}

\begin{figure}[t]
  \centering
  \subfigure[515 Hz]{\includegraphics[scale = .5, trim=250 400 250 370, clip = true]{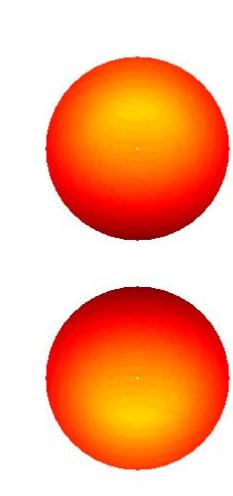}} \quad \quad %
 \subfigure[3.63 kHz]{\includegraphics[scale = .5, trim=250 295 250 290, clip = true]{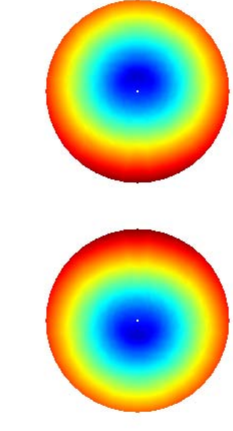}}
\caption{Two wires example of Sec.~\ref{sec:2cond} in the case of a separation of 25 mm: current density plot for two different frequencies, obtained with MoM-SO.}
\label{fig:current}
\end{figure}

The p.u.l. parameters have been computed from 1~Hz to 1~MHz with the truncation order $N_p$ set to 3. No noticeable changes have been observed beyond this value. The computation of the parameters took 15~ms per frequency sample on a 3.4 GHz CPU. The discretization of the Green's function which leads to the matrix $\matr{G}$ took less than 5~ms. Figures~\ref{fig:res1} and~\ref{fig:ind1} compare the p.u.l. parameters computed with MoM-SO against the results obtained from the two analytic formulas. In all cases, the numerical results correctly approach the exact high frequency value given by formulas~\pref{eq:R} and \pref{eq:Lext}. In the case of wide separation, shown in the top panels, the numerical results also correctly predict the frequency-dependent behavior of the wire's internal impedance~\pref{eq:Zint} due to skin effect. A little discrepancy is visible in the inductance at low frequency (top panel of Fig.~\ref{fig:ind1}), due to the small but not negligible proximity effect. The error introduced by~\pref{eq:Zint} becomes more significant when the wires separation is reduced to 25 mm, as shown in the bottom panel of Figures~\ref{fig:res1} and~\ref{fig:ind1}. The non-uniform current distribution induced  by the wires proximity is visible in Fig.~\ref{fig:current},
which also shows the development of skin effect. 

\subsection{Three-Phase Armored Cable}
\label{sec:cable}

\begin{figure}[t]
\centering
\includegraphics[width=.6\columnwidth, viewport= 100 220 520 600]{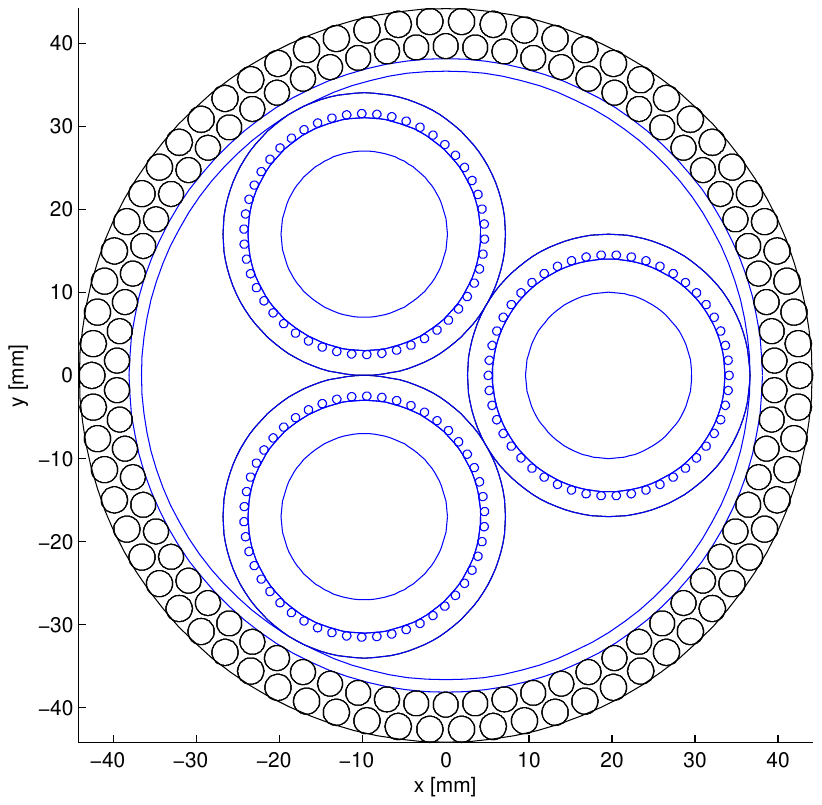}
\caption{Cross section of the three-phase cable with steel armoring and wire sheath considered in Sec.~\ref{sec:cable}}
\label{fig:Geometry2}
\end{figure}

We consider the three-phase armored cable in Fig.~\ref{fig:Geometry2} which features three wire screens and a steel armoring, for a total of $293$ circular subconductors. The key parameters are listed in Tables~\ref{tab:cabledata} and~\ref{tab:armordata}, respectively. Using MoM-SO we computed the $3\times 3$ series impedance matrix with respect to the three phase conductors with the screens continuously grounded along the cable.

\renewcommand{\arraystretch}{1.2}

\begin{table}[t]
\centering
\caption{Characteristics of the cables in the example of Sec.~\ref{sec:cable}.}
\begin{tabular}{|c|c|}
\hline
 {\bf Item} & {\bf Parameters} \\ \hline
Core & $\sigma=58 \cdot 10^6 \, {\rm S/m}$, $r=10.0 \, {\rm mm}$ \\ \hline
Insulation & $t=4.0 \, {\rm mm}$, $\epsilon_r=2.3 $\\ \hline
Wire screen & 32 wires, $r=0.5 \, {\rm mm}$, $\sigma=58 \cdot 10^6 \, {\rm S/m}$ \\ \hline
Jacket & $t=2 \, {\rm mm}$, $\epsilon_r=2.3$ \\ \hline
\end{tabular}
\label{tab:cabledata}
\end{table}

\begin{table}[t]
\centering
\caption{Armor characteristics for the structure considered in Sec.~\ref{sec:cable}.}
\begin{tabular}{|c|c|}
\hline
 {\bf Item} & {\bf Parameters} \\ \hline
Armor outer diameter & 88.26 mm  \\ \hline
Wire diameter & 3 mm \\ \hline
Conductivity & $10^7$ S/m \\ \hline
$\mu_r$ & 100 \\ \hline
N.o. wires per layer & 70 \\ \hline
\end{tabular}
\label{tab:armordata}
\end{table}

Table~\ref{tab:impedance} shows the calculated positive and zero sequence resistance and reactance per km. The computation has been performed with three different truncation orders: $N_p=0$, $N_p=3$ and $N_p=7$. As a validation we used a FEM implementation \cite{B09} with a very fine mesh (177,456 triangles). It is observed that with orders $N_p=3$ and $N_p=7$ we get a result which deviates by less than 1\% from the FEM result. 

\begin{figure}[t]
\centering
\includegraphics[width = .75\columnwidth, viewport= 130 225 510 550]{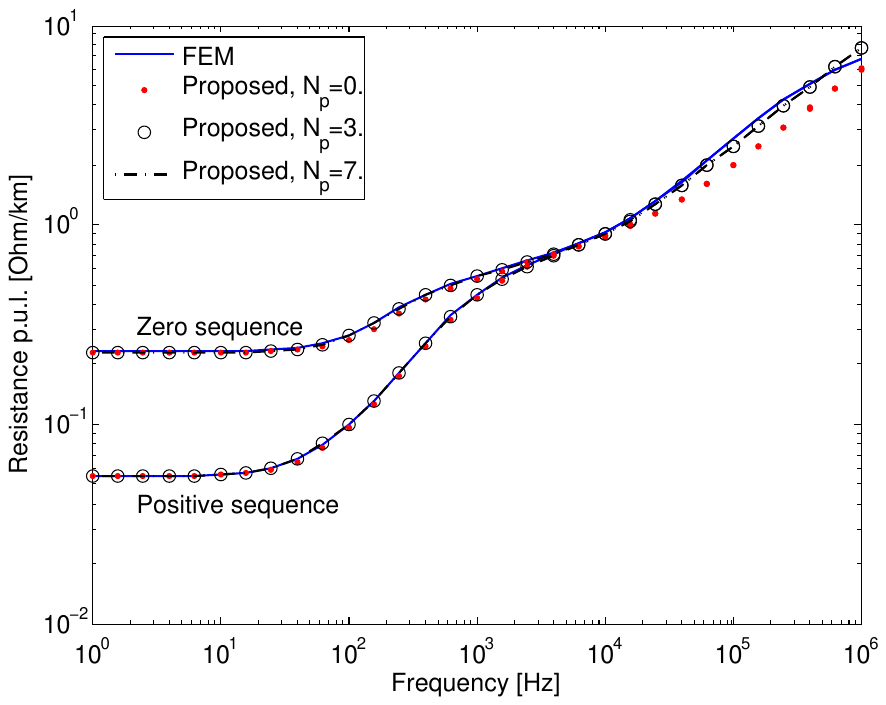}
\caption{P.u.l. resistance of the three-phase cable of Sec.~\ref{sec:cable}, obtained with MoM-SO and FEM. For MoM-SO, three different truncation orders $N_p$ are considered.}
\label{fig:Resistance2}
\end{figure}

\begin{figure}[t]
\centering
\includegraphics[width=.75\columnwidth, viewport= 130 225 510 550]{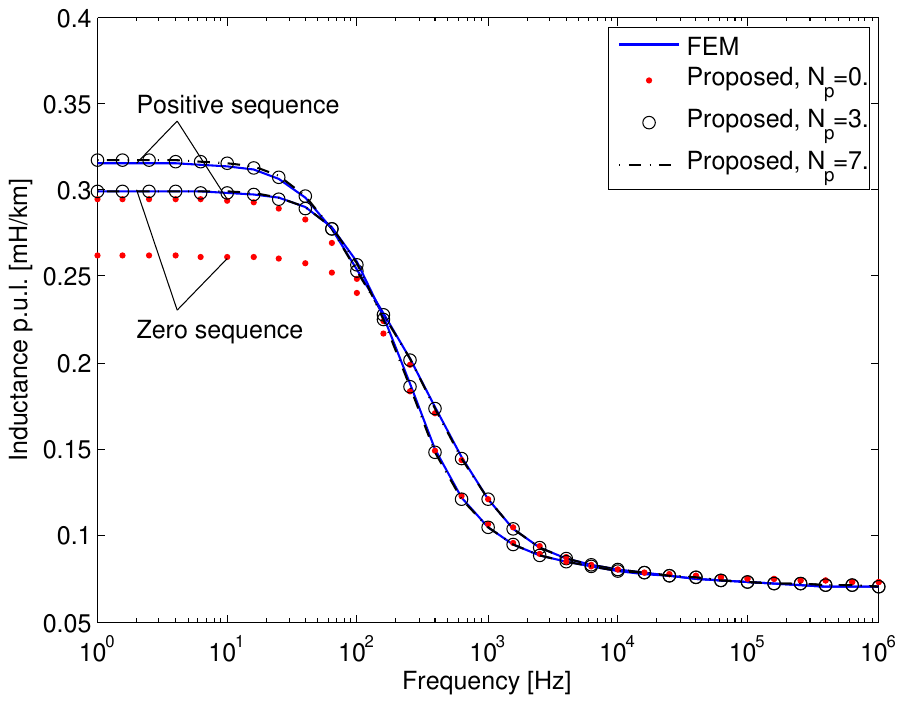}
\caption{P.u.l. inductance of the three-phase cable of Sec.~\ref{sec:cable}, obtained with MoM-SO and FEM. For MoM-SO, three different truncation orders $N_p$ are considered.}
\label{fig:Inductance2}
\end{figure}

Figs.~\ref{fig:Resistance2}  and \ref{fig:Inductance2} show the positive and zero sequence resistance and inductance as a function of frequency, from $1\, {\rm Hz}$ to $1\,{\rm MHz}$. It is observed that with $N_p=0$, significant errors result as the proximity effects are ignored. Indeed, by setting $N_p=0$ in~\pref{eq:Eseries} and~\pref{eq:Jseries}, one assumes a circularly-symmetric current distribution on the conductors. With $N_p=3$ and $N_p=7$, a virtually identical result is achieved which agrees very well with the FEM result. At very high frequencies, however, the FEM result deviates somewhat from that of the proposed approach since the mesh division is not sufficiently fine to properly account for the very small skin depth. In the proposed technique, instead, skin effect is implicitly and fully described by the surface admittance operator, and does not affect the discretization of the problem, which depends only on the proximity of the conductors. As a result, the level of discretization, controlled by $N_p$, does not have to be increased as frequency grows, making MoM-SO much more efficient than FEM.

Timing results, reported in Table~\ref{tab:Timing}, demonstrate the excellent performance of the developed algorithm. With MoM-SO, there is first a computation time for the Green's matrix~$\matr{G}$ of 11.6~s ($N_p=3$) or 16.5~s ($N_p=7$). Matrix $\matr{G}$ has to be evaluated only once, since it does not depend on frequency. Then, for computing each frequency sample one needs 2.01~s for $N_p=3$ or 15.5~s for $N_p=7$. Since $N_p=3$ was found sufficient for obtaining accurate results, the total computational cost for computing the 31 samples in this example is $T=11.6+31\times2.01=73.9~s$. The computation time using FEM is much higher, requiring 440~s per frequency sample. This is 220 times slower than the per-sample computation time of 2.01~s using the new method with $N_p=3$. We can therefore safely state that MoM-SO is at least 100 times faster than the FEM approach when several frequency samples are needed.      

\begin{table}[t]
\centering
\caption{Positive- and zero-sequence impedance of the three-phase cable of Sec.~\ref{sec:cable} at $50 \, {\rm Hz}.$ MoM-SO is compared against a finite element approach~\cite{B09}.}
\begin{tabular}{| c | c | c | c | c |}
\hline
& \multicolumn{3}{|c|}{\bf MoM-SO (proposed)} & \\
\cline{2-4}
& ${N_p=0}$ & $ N_p=3$  & $ N_p=7$ & \bf{FEM} \\ \hline
$R_{+}[\Omega/{\rm km}]$  & 0.06905 & 0.07259 & 0.07261 & 0.07218 \\ \hline
$X_{+}[\Omega/{\rm km}]$  & 0.08703 & 0.09042	& 0.09048  & 0.09041 \\ \hline
$R_{0}[\Omega/{\rm km}]$  & 0.2386	& 0.2437  & 0.2438  & 0.2459 \\ \hline
$X_{0}[\Omega/{\rm km}]$  & 0.08033 & 0.08943 & 0.08958  & 0.08975 \\ \hline
\end{tabular}
\label{tab:impedance}
\end{table}

\begin{table}[t]
\centering
\caption{Three-phase cable example of Sec.~\ref{sec:cable}: computation time for MoM-SO and FEM.}
\begin{tabular}{| p{2.0cm} | c | c | c | c |}
\hline
& \multicolumn{3}{|c|}{\bf MoM-SO (proposed)} & \\
\cline{2-4}
& ${ N_p=0}$ & ${N_p=3}$  &  ${N_p=7}$ & {\bf FEM} \\ \hline
Green's  function
discretization & 11.6~s & 13.7~s & 16.5~s &  \\ \hline
Impedance computation (per frequency sample) & 0.085~s & 2.01~s & 15.5~s & 440~s* \\ \hline
\end{tabular}
\label{tab:Timing}
\newline
\newline
{All computations were performed on a system\\ with a 2.5 GHz CPU and 16~GB of memory.\\
*Mesh size: 177,456 triangles.}
\end{table}

\section{Discussion}

\subsection{Computational Cost}
\label{sec:cost}

A few remarks on the computational cost of MoM-SO are in order. The most expensive step in evaluating~\pref{eq:ppulpR} and~\pref{eq:ppulpL} is the LU factorization of the matrix
\begin{equation}
\matr{M} = \matr{1} - j\omega \mu_0 \matr{Y}_s \matr{G}\,,
\label{eq:M}
\end{equation}
which is used to compute the term $(\matr{1} - j\omega \mu_0 \matr{Y}_s \matr{G})^{-1} \matr{Y}_s \matr{U}$. The matrix $\matr{M}$ has size $N \times N$, where $N$ is the total number of unknowns used to discretize the problem~\pref{eq:N}. If we let $N_p = 3$ for all conductors, we obtain that $N = 7P$. Therefore, the number of unknowns $N$ scales well with the number of conductors $P$, and it remains moderate even in presence of hundreds of conductors. In the example of Sec.~\ref{sec:cable}, which has 293 conductors, MoM-SO uses $N = 2051$ unknowns, as opposed to the 177,464 unknowns required by FEM. Even if the MoM-SO matrix~\pref{eq:M} is full while the FEM matrix is very sparse, the huge difference in size makes MoM-SO faster than FEM, as shown by the numerical results. The remarkable saving of unknowns stems from the use of a surface formulation instead of a volume formulation, where one must mesh the entire volume of the conductors and, possibly, also of the surrounding medium. The FEM code~\cite{B09} used in this paper is an in-house program which adapts state-of-the-art routines in Matlab's PDE Toolbox to the Weiss-Scendes one-step FEM method~\cite{Wei82} for series impedance computation. Although the usage of a different FEM implementation or a different meshing strategy may improve the computational efficiency, the need for a very large number of triangles cannot be overcome. When computing frequency samples over a wide frequency band, the mesh must have a fine resolution over the entire solution domain to capture the low frequency behavior, and at the same time have a very fine resolution at the conductor surfaces to capture the pronounced skin effect at high frequency. The MoM-SO approach fundamentally overcomes this issues, since it does not require any meshing of the cross section.

\subsection{Relation with Existing EMTP Tools}

In the numerical example of Sec.~\ref{sec:cable}, we considered a commonly applied three-phase cable design which features wire screens and a stranded steel armoring. This cable was modeled with MoM-SO with an explicit representation of each strand. In available EMTP-tools, one would have to model the screens and the armor by equivalent tubular conductors. Such approach leads to very fast computations but errors are inevitably introduced, in particular for the steel armoring where the air gaps between magnetic strands cannot be easily accounted for by an equivalent tubular conductor. We have shown that with MoM-SO such cables can be modeled in full detail with an acceptable CPU time, while taking into account both skin and proximity effects in every phase conductor, and in every wire and armor strand.

\subsection{Effect of Lossy Ground}

The current version of MoM-SO  does not permit to include the effect of a lossy ground. However, in the case of transients involving armored and pipe-type cables, the effect of the ground return is often small and can be ignored, at least when the transient effect does not include current injection to ground. The authors are currently extending the method to include ground return effects and tubular conductors (sheaths). As for the example of the three-phase cable in Sec.~\ref{sec:cable}, the modeling in this paper is fully applicable since the sheath conductor consists of wires and because the thick armor permits to ignore the external medium.

\section{Conclusion}
We presented an efficient algorithm for computing the series impedance of systems of round conductors. The method combines a surface operator with the method of moments which permits to compute the complete series impedance matrix while taking into account both skin and proximity effects. This capability is of major importance in cable system modeling as the short lateral distance between cables often leads to significant proximity effects. Due to its efficient discretization of the underlying electromagnetic problem, the algorithm outperforms state-of-the-art techniques based on finite elements by a factor of about 100. The computed resistance and inductance can be used for an accurate prediction of electromagnetic transients in EMTP-type programs when combined with an appropriate frequency-dependent cable model.

\appendix
\section*{Analytical evaluation of the Green's matrix~$\matr{G}$}
\label{App:Green}

The discretization of the Green's function requires the computation of the double integral~\pref{eq:greenentry}. After substitution of~\pref{eq:green}, the integral reads
\begin{equation}
\matr{G}^{(p,q)}_{n'n}=\frac{1}{(2 \pi)^2} \int_{0}^{2\pi} f_n(\theta) e^{-j n' \theta} d\theta \,.
\label{eq:greenentry2}
\end{equation}
where
\begin{equation}
f_n(\theta) = \frac{1}{2\pi} \int_{0}^{2\pi} 
 \ln \abs{\vec{r}_p(\theta)-\vec{r}_q(\theta')} e^{j n \theta'} d\theta'\,.
\label{eq:fn}
\end{equation}
We first calculate the integral in~\pref{eq:fn}, which can be expanded using~\pref{eq:contour} to obtain
\begin{align}
f_n(\theta) & = \frac{1}{2\pi} \int_0^{2\pi} \ln \abs{ \vec{r''}(\theta) - a_q (\cos \theta' \ver{x} 
+ \sin \theta' \ver{y} )} e^{j n \theta'} d \theta' =  \nonumber \\
&\frac{1}{4\pi} \int_0^{2\pi} \ln  \left [ (r'')^2  +a_q^2 - 2 a_q r'' \cos (\theta''-\theta') \right ] e^{j n \theta'} d \theta'\,,
\label{eq:fnexp}
\end{align}
where $\vec{r''}(\theta)$ is the auxiliary vector
\begin{equation}
\vec{r''}(\theta) = \vec{r}_p(\theta) - (x_q \ver{x} + y_q \ver{y} )= r'' ( \cos \theta'' \ver{x} + \sin \theta'' \ver{y})\,,
\label{eq:r2}
\end{equation}
which is constant with respect to the integration variable $\theta'$. We denote its modulus and its angle with $r''$ and $\theta''$ respectively\footnote{For the sake of clarity of the notation, we omit from $r''$ and $\theta''$ the dependence on $\theta$.}. The solution to the last integral in~\pref{eq:fnexp} is presented in \cite{Pau07} and reads
\begin{equation}
f_n(\theta) = \begin{cases}
\ln(r'')	& \text{for } n=0\,,\\
-\frac{a_q^{\abs{n}} e^{j n \theta''}}{2 \abs{n} (r'')^{\abs{n}}} & \text{for } n \neq 0\,.
\end{cases}
\label{eq:fn2}
\end{equation}
Next, we solve the integral in~\pref{eq:greenentry2}. The solution involves several cases, which are itemize for better readability:
\begin{itemize}
\item {$ p \neq q$ , $n=0$:} in this case, the integral in~\pref{eq:greenentry2} is analogous to  \pref{eq:fn} and can be solved with the formulas given in~\cite{Pau07}. When $n' =0$, we have
\begin{equation}
\matr{G}^{(p,q)}_{0,0} = 
\frac{1}{2\pi} \ln (d_{p,q})\,,
\label{eq:Gblock8a}
\end{equation}
and when $n' \neq 0$ we have
\begin{equation}
\matr{G}^{(p,q)}_{n',0} = 
-\frac{1}{4\pi \abs{n'}} \left ( \frac{a_p}{d_{p,q}} \right )^{\abs{n'}} \left ( -\frac{x_{p,q}-j y_{p,q}}{d_{p,q}} \right )^{n'}\,,
\label{eq:Gblock8b}
\end{equation}
where $x_{p,q} = (x_p - {x_q})$, $y_{p,q} = ({y_p} - {y_q})$, and 
\begin{equation}
	d_{p,q} = \sqrt{x_{p,q}^2 + y_{p,q}^2}\,.
\end{equation}
\item {$ p \neq q$ , $n>0$, $n' \ge 1$:} if we manipulate the integrand function as follows
\begin{multline}
f_n(\theta) = -\frac{(a_q)^n e^{j n \theta''}}{2 n (r'')^{n}}=
-\frac{(a_q)^n}{2n} \left [ \frac{r'' e^{j \theta''}}{(r'')^2} \right ]^n = \nonumber \\
 -\frac{(a_q)^n}{2n} \left [ \frac{x_{p,q} + a_p \cos \theta + j (y_{p,q} + a_p \sin \theta)}{(x_{p,q} + a_p \cos \theta)^2 + (y_{p,q} + a_p \sin \theta)^2} \right ]^n = \nonumber \\
-\frac{(a_q)^n}{2n} \frac{1}{(x_{p,q} - j y_{p,q} + a_p e^{-j\theta})^n}\,,
\label{eq:fn3}
\end{multline}
we can rewrite~\pref{eq:greenentry2} as a complex integral 
\begin{equation}
\matr{G}^{(p,q)}_{n',n} = -\frac{j (a_q)^n}{8 \pi^2 n (a_p)^{n'}} \ointclockwise \frac{z^{n'-1}}{(x_{p,q} -j y_{p,q} + z)^n} dz\,,
\label{eq:Gblock4}
\end{equation}
performed over the closed path $z = a_pe^{-j\theta}$ with $\theta = [0, 2\pi]$. Since we assume $n' \ge 1$, the integrand function in~\pref{eq:Gblock4} will have no poles inside the integration path and, using the residue theorem~\cite{Zil06}, we have that
\begin{equation}
\matr{G}^{(p,q)}_{n',n} = 0\,.
\label{eq:Gblock5}
\end{equation}
\item {$ p \neq q$ , $n>0$, $n' < 1$:} when $n' < 1$, the integrand function in~\pref{eq:Gblock4} has a pole in $z = -x_{p,q} + j y_{p,q}$. Applying again the residue theorem we obtain
\begin{align}
\matr{G}^{(p,q)}_{n',n} = \frac{-\pi(a_q)^n}{(-a_p)^{n'}} \binom{n-n'-1}{-n'}
\frac{(d_x-jd_y)^{-n+n'}}{(2\pi)^2 n}\,,
\label{eq:Gblock6}
\end{align}
with $\binom{n}{m}$ being the binomial coefficient.
\item {$ p \neq q$ , $n<0$:} the Green's matrix entries for $n <0$ can be obtained from~\pref{eq:Gblock5} and~\pref{eq:Gblock6} by symmetry
\begin{align}
\matr{G}^{(p,q)}_{n',n} = (\matr{G}^{p,q}_{-n',-n})^*\,,
\end{align}
where $^*$ denotes complex conjugation.
This identity follows from the symmetry relation $f_{-n}(\theta) = f_n(\theta)^*$.
\item {$ p = q$:} in this case we have that $x_{p,q} = y_{p,q} = 0$, and using \pref{eq:Gblock4} one can easily show that
\begin{equation}
\matr{G}^{(p,p)}_{n',0} = 
\begin{cases}
\frac{1}{2\pi} \ln a_p & \text{if } n' = 0\,,\\
0 & \text{if } n' \neq 0\,,
\end{cases}
\label{eq:Gblock10}
\end{equation}
\begin{equation}
\matr{G}^{(p,p)}_{n',n} = 
\begin{cases}
-\frac{1}{4 \pi \abs{n}} & \text{if } n \neq 0, n'=n\,,\\
0 & \text{if } n \neq 0 , n' \neq n\,.
\end{cases}
\label{eq:Gblock9}
\end{equation}
\end{itemize}

\bibliography{biblio}
\bibliographystyle{IEEEtran}

\end{document}